\documentclass[a4paper]{article}           
\usepackage[latin1]{inputenc}     
\usepackage[T1]{fontenc}          
\usepackage[dvips]{graphicx}      
\usepackage{fancybox}		   
\usepackage{makeidx}              
\usepackage{amsmath}
\usepackage{amssymb}

\title{Volatility Swap Under The SABR Model}     
\author{Simon Bossoney}        

\makeindex		    
\begin{document}                  

\maketitle                        
\newpage
\tableofcontents                  
\newpage

\begin{abstract}
The SABR model is shortly presented and the volatility swap explained. The fair value for a volatility swap is then computed using the usual theory in financial mathematics. An analytical solution using confluent hypergeometric functions is found. The solution is then verified using Rama Cont's functional calculus.
\end{abstract}
\section{Introduction}

Mathematical models for the computation of financial derivatives have gained some importance in the financial industry. A breakthrough was realized with the Black \& Scholes model. Though widely applied, this model soon began to show its limits and shortcomings. Among other deficiencies, the hypothesis of constant volatility for the underlying appeared as a much to strong assumption. It is usually now dealt with by quoting a so-called implied volatility, which, when inserted in the Black \& Scholes equation for the option pricing, gives the traded prices for the latter. 

\subsection{The SABR Model}            

As another Ansatz, models with time-dependent volatilities or even stochastic volatilities where introduced. A famous example is the Heston-model, where the variance of the underlying is itself stochastic. Another widely used model is the SABR-model, where this time the volatility is stochastic. Under the forward measure, the forward prices $F_t$ for an underlying is governed by the stochastic process
\begin{gather}
dF_t=\sigma_tF_tdW_t,\nonumber\\
d\sigma_t=\alpha\sigma_tdZ_t,
\end{gather}
where $\alpha$ is some positive constant and where $W_t$ and $Z_t$ are Wiener processes with a correlation coefficient $-1\leq\rho\leq1$ \cite{hagan}. In general, the first stochastic differential equation reads $dF_t=\sigma_tF_t^\beta dW_t$, with $0\leq\beta\leq1$, but we are going to work with $\beta=1$ throughout our computations. This will have no influence in the following, since in our definition of the volatility swap, only the stochastic process governing the volatility will be considered.
European call options are then priced under the forward measure as 
\begin{gather}
S(F,\sigma,K,t)=p(t,T)\mathbb{E}\big((F_T-K)^+\vert F_t=F,\sigma_t=\sigma\big),
\end{gather}
where $p(t,T)$ is the price at $t$ of a zero-coupon bond paying $1$ at maturity $T$ (see \cite{bjoerk09}).

\section{Volatility Swaps}

As it became obvious that volatilities and variances for various underlyings behave in a stochastic manner on the market, several derivative instruments dealing with those quantities  where introduced. 

\subsection{Principle of The swaps}

A variance swap consists of a strike $K$ and a realized average variance $1/T\int_{t_o}^{t_0+T}\sigma_t^2dt$ over some time-period $t_0,t_0+T$. The pay-off function is then the difference of these two quantities. There are known closed form solution for variance swaps, but it seems, that these instruments are rather volatile and risky. \\
A volatility swap is a bet on a realized averaged volatility $1/T\sqrt{\int_{t_o}^{t_0+T}\sigma_t^2dt}$ over some time-period $t_0,t_0+T$. By taking the square root of the realized variance, this instrument has a somewhat less volatile behavior than a variance swap. But mathematically, it is exactly this square root that makes computations more difficult.\\
Here we propose to compute exact solutions to the fair value of a volatility swap
$$p(t,T)\Big(\mathbb{E}\big(\frac{1}{T}\sqrt{\int_{t_o}^{t_0+T}\sigma_t^2dt}\vert\sigma_t=\sigma\big)-K\Big),$$
under the SABR model with $t\in[t_0,t_0+T]$. In the following, we shall write 
$$\kappa_t:=\mathbb{E}\big(\frac{1}{T}\sqrt{\int_{t_o}^{t_0+T}\sigma_t^2dt}\vert\sigma_t=\sigma\big),$$ so that the fair price of a volatility swap becomes $p(t,T)(\kappa_t-K)$. The task is now to compute $\kappa_t$.

\subsection{Getting Rid of The Square Root}

It is a well-known fact, that for $y\in\mathbb{R}^+$
\begin{equation}
\sqrt{y}=\frac{1}{\sqrt{\pi}}\int_0^\infty\frac{1-\exp(-x^2y)}{x^2}dx
\end{equation}
we therefore obtain
\begin{eqnarray}
\kappa_t&=&\mathbb{E}\big(\frac{1}{T}\sqrt{\int_{t_o}^{t_0+T}\sigma_t^2dt}\vert\sigma_t=\sigma\big)\nonumber\\
&=&\frac{1}{T\sqrt{\pi}}\mathbb{E}\big(\int_0^\infty\frac{1-\exp(-x^2\int_{t_o}^{t_0+T}\sigma_t^2dt)}{x^2}\big\vert\sigma_t=\sigma\big)\nonumber\\
&=&\frac{1}{T\sqrt{\pi}}\int_0^\infty\frac{1-\mathbb{E}\big(\exp(-x^2\int_{t_o}^{t_0+T}\sigma_t^2dt)\big\vert\sigma_t=\sigma\big)}{x^2}.
\end{eqnarray}
We will hence have to compute the quantity

\begin{equation}
f(t,\sigma,x):=\mathbb{E}\big(\exp(-x^2\int_{t_o}^{t_0+T}\sigma_t^2dt)\big\vert\sigma_t=\sigma\big)
\end{equation}
and integrate along $x$ as in the previous equality.\\
Note, that $x$ is not stochastic but a variable of integration. Also, we did not fully justify the interchange of the order of integration, which shall be justified a posteriori. Also, if $t_0\leq t\leq t_0+T$, then $f$ will also depend on the realized volatility $\nu_t:=\int_{t_0}^t\sigma_t^2dt$ "up to $t$". We suppose at first, that $t_0\leq t\leq t_0+T$ and write

\begin{eqnarray}
f(t,\sigma,\nu_t,x)&:=&\mathbb{E}\big(\exp(-x^2\int_{t_o}^{t_0+T}\sigma_t^2dt)\big\vert\sigma_t=\sigma\big)\nonumber\\
&=&\mathbb{E}\big(\exp(-x^2\nu_t)\exp(-x^2\int_{t}^{t_0+T}\sigma_t^2dt)\big\vert\sigma_t=\sigma\big)\nonumber\\
&=&\exp(-x^2\nu_t)\mathbb{E}\big(\exp(-x^2\int_{t}^{t_0+T}\sigma_t^2dt)\big\vert\sigma_t=\sigma\big)\nonumber\\
&\equiv&\exp(-x^2\nu_t)\varphi(t,\sigma,x).
\end{eqnarray}

\subsection{A differential Equation for $\varphi$}

The Fenman-Kac formula gives the correspondence between expectation computations for a stochastic process and differential equations. For a stochastic process governed by the equation
\begin{gather}
dX_t=\mu(X_t,t)dt+\sigma(X_t,t)dW_t,
\end{gather}

the function

\begin{gather}
u(x,t):=\mathbb{E}\big(\int_t^T\exp(-\int_t^sV(X_\tau)d\tau)f(X_s,s)ds+\exp(-\int_t^TV(X_s)ds)\Psi(X_T)\big)
\end{gather}

is the solution to the differential equation
\begin{gather}
\partial_tu(x,t)+\mu(x,t)\partial_xu(x,t)+\frac{\sigma^2(x,t)}{2}\partial^2_xu(x,t)-V(x,t)u(x,t)+f(x,t)=0,
\end{gather}
subject to the terminal condition
\begin{gather}
u(x,T)=\Psi(x).
\end{gather}
In our situation, we have $\mu(\sigma_t,t)=0=f(\sigma_t,t)$, $\Psi(x)=1$, $V(\sigma_t)=x^2\sigma_t^2$ and the stochastic diffusion term is $\alpha\sigma_t$, so that $\varphi(t,\sigma,x)$ is, for a given $x$, solution to the differential equation
\begin{gather}
\partial_t\varphi(t,\sigma,x)+\frac{\alpha^2\sigma^2}{2}\partial^2_\sigma\varphi(t,\sigma,x)-x^2\sigma^2\varphi(t,\sigma,x)=0\\
\Leftrightarrow-\frac{2}{\alpha^2}\partial_t\varphi(t,\sigma,x)=\sigma^2\partial^2_\sigma\varphi(t,\sigma,x)-\frac{2x^2\sigma^2}{\alpha^2}\varphi(t,\sigma,x),
\end{gather}
subject to the terminal condition
\begin{gather}
\varphi(T,\sigma,x)=1.
\end{gather}

Note, that here, $x$ is only a parameter, over which we will have to integrate later on. Here, $x$ is indexing a family of differential equations. We shall hence drop it from the variables in $\varphi$. Writing $y:=\sqrt{2}x\sigma/\alpha$ and $\psi(t,y):=\varphi(t,\sigma(y))$, 
we are to solve
\begin{gather}
-\frac{2}{\alpha^2}\partial_t\psi(t,y)=y^2\partial^2_y\psi(t,y)-y^2\psi(t,y),\label{diffEq10}
\end{gather}
subject to the terminal condition
\begin{gather}
\psi(T,y)=1.
\end{gather}

Clearly, we will have $\varphi(t,\sigma,x)=\psi(t,\frac{\sqrt{2}x\sigma}{\alpha})$.

\section{Solving the differential equation}

If it wasn't for the missing $y\partial_y$ term, the differential operator appearing on the right-hand side of equation  \ref{diffEq10} would be the modified Bessel differential operator, whose solutions are well-known. In order to retrieve this equation, we define $g(t,y):=y^{-1/2}\psi(t,y)$. This function satisfies the equation

\begin{gather}
\frac{1}{4}g(t,y)-\frac{2}{\alpha^2}\partial_tg(t,y)=y^2\partial^2_yg(t,y)+y\partial_yg(t,y)-y^2g(t,y),\label{diffEq11}
\end{gather}
subject to the terminal condition
\begin{gather}
g(T,y)=\frac{1}{\sqrt{y}}.
\end{gather}

The solutions to the equation
\begin{gather}
k^2f(y)=y^2\partial^2_yf(y)+y\partial_yf(y)-y^2f(y),
\end{gather}
are the modified Bessel functions of the first and second kind $I_k(y)$ and $K_k(y)$. The functions $I_k(y)$ read
\begin{gather}
I_k(y)=\sum_{m\geq0}\frac{1}{m!\Gamma(k+m+1)}\Big(\frac{y}{2}\Big)^{2m+k}.
\end{gather}
The terminal condition $y^{-1/2}$ can be expressed in terms of these functions as
\begin{gather}
\frac{1}{\sqrt{y}}=\frac{1}{\sqrt{2}}\sum_{n\geq0}\frac{(-1)^n(2n-1/2)\Gamma(n-1/2)}{n!}I_{2n-1/2}(y).
\end{gather}
The convergence of this series is pointwise. We may now write the solution to the equation \ref{diffEq11} as
\begin{gather}
g(t,y)=\sum_{n\geq0}\frac{(-1)^n(2n-1/2)\Gamma(n-1/2)}{\sqrt{2}n!}I_{2n-1/2}(y)\exp\big(E_n(T+t_0-t)\big).\label{sol1}
\end{gather}
where $E_n:=\frac{\alpha^2}{2}\big((2n-1/2)^2-1/4\big)=\alpha^2n(2n-1)$. Consequently,
\begin{gather}
\psi(t,y)=\sum_{n\geq0}\frac{(-1)^n(2n-1/2)\Gamma(n-1/2)}{n!}\big(\frac{y}{2}\big)^{1/2}I_{2n-1/2}(y)\exp\big(E_n(T+t_0-t)\big).\label{sol2}
\end{gather}

\section{Recovering The Square Root}

We shall now compute $\kappa_t$ for a volatility swap under the SABR model, by integrating over $x$. At first, we will suppose $t_0\leq t\leq t_0+T$. The case $t<t_0$ will be treated afterwards. We thus have

\begin{eqnarray}
\kappa_t&=&\frac{1}{T\sqrt{\pi}}\int_0^\infty\frac{1-\exp(-x^2\nu_t)\varphi(t,\sigma,x)}{x^2}dx\nonumber\\
&=&\frac{1}{T\sqrt{\pi}}\int_0^\infty\frac{1-\exp(-x^2\nu_t)+\exp(-x^2\nu_t)\big(1-\varphi(t,\sigma,x)\big)}{x^2}dx\nonumber\\
&=&\frac{1}{T}\Big\{\sqrt{\nu_t}+\frac{1}{\sqrt{\pi}}\int_0^\infty\frac{\exp(-x^2\nu_t)\big(1-\varphi(t,\sigma,x)\big)}{x^2}dx\Big\}\nonumber\\
&=&\frac{1}{T}\Big\{\sqrt{\nu_t}+\frac{1}{\sqrt{\pi}}\int_0^\infty\frac{\exp(-x^2\nu_t)\big(1-\psi(t,\frac{\sqrt{2}x\sigma}{\alpha})\big)}{x^2}dx\Big\}\nonumber\\
&=&\frac{1}{T}\Big\{\sqrt{\nu_t}+\frac{\sqrt{2}\sigma}{\alpha\sqrt{\pi}}\int_0^\infty\frac{\exp(-\frac{\alpha^2\nu_t}{2\sigma^2}y^2)\big(1-\psi(t,y)\big)}{y^2}dy\Big\}\nonumber
\end{eqnarray}

To lighten up the notations, we shall write $J:=\int_0^\infty\frac{\exp(-\frac{\alpha^2\nu_t}{2\sigma^2}y^2)\big(1-\psi(t,y)\big)}{y^2}dy$. Moreover, we write $a_n:=\frac{(-1)^n(2n-1/2)\Gamma(n-1/2)}{n!}$ and $F_n(y):=\big(\frac{y}{2}\big)^{1/2}I_{2n-1/2}(y)$. We will split the integral as

\begin{eqnarray}
J&=&\int_0^\infty\frac{\exp(-\frac{\alpha^2\nu_t}{2\sigma^2}y^2)\big(1-\psi(t,y)\big)}{y^2}dy\,\,\,\,\,\,\,\,\,\,\,\,\,\text{($z:=\frac{2\sigma^2}{\alpha^2\nu_t}$)}\nonumber\\
&=&\int_0^\infty\frac{\exp(-\frac{y^2}{z})\big(1-\sum_{n\geq0}a_nF_n(y)\exp(E_n(T+t_0-t))\big)}{y^2}dy\nonumber\\
&=&J_0+J_\infty,\nonumber
\end{eqnarray}
where
\begin{eqnarray}
J_0&=&\int_0^\infty\frac{\exp(-\frac{y^2}{z})\big(1-a_0F_0(y)\big)}{y^2}dy\,\,\,\text{ and }\nonumber\\
J_\infty&=&-\int_0^\infty\frac{\exp(-\frac{y^2}{z})\big(\sum_{n\geq1}a_nF_n(y)\exp(E_n(T+t_0-t))\big)}{y^2}dy.\nonumber
\end{eqnarray}
Observe, that $F_n(y):=\sum_{m\geq0}\frac{1}{m!\Gamma(m+2n+1/2)}\big(\frac{y}{2}\big)^{2m+2n}$. To compute $J_0$, we insert $a_0=-1/2\Gamma(-1/2)=\sqrt{\pi}$, and $F_0=\sum_{m\geq0}\frac{1}{m!\Gamma(m+1/2)}\big(\frac{y}{2}\big)^{2m}$. Therefore,
\begin{eqnarray}
J_0&=&\int_0^\infty\frac{\exp(-\frac{y^2}{z})\big(1-\sqrt{\pi}F_0(y)\big)}{y^2}dy\nonumber\\
&=&\int_0^\infty\frac{\exp(-\frac{y^2}{z})\sum_{m\geq1}\frac{-\Gamma(1/2)}{m!\Gamma(m+1/2)}\big(\frac{y}{2}\big)^{2m}}{y^2}dy\nonumber\\
&=&\int_0^\infty\exp(-\frac{y^2}{z})\sum_{m\geq1}\frac{-\Gamma(1/2)}{m!\Gamma(m+1/2)}\big(\frac{y}{2}\big)^{2m}\frac{1}{4}\big(\frac{y}{2}\big)^{-2}dy\nonumber\\
&=&\sum_{m\geq1}\frac{-\Gamma(1/2)}{4m!\Gamma(m+1/2)}\int_0^\infty\exp(-\frac{y^2}{z})\big(\frac{y}{2}\big)^{2m-2}dy,\nonumber
\end{eqnarray}
where the interchange of the integral and the sum is justified by the fact, that all terms which are to integrate have the same sign, so that the convergence of the integral of the sum is equivalent to the convergence of the sum of the integrals.\\
This integral is fairly easily computed:
\begin{eqnarray}
\int_0^\infty\exp(-\frac{y^2}{z})\big(\frac{y}{2}\big)^{2m-2}dy&=&\int_0^\infty\exp(-s)\big(\frac{sz}{4}\big)^{m-1}\big(\frac{z}{4s}\big)^{1/2}ds\nonumber\\
&=&\int_0^\infty\exp(-s)\big(\frac{z}{4}\big)^{m-1/2}s^{m-3/2}ds\nonumber\\
&=&\big(\frac{z}{4}\big)^{m-1/2}\Gamma(m-1/2).\nonumber
\end{eqnarray}
We arrive at
\begin{eqnarray}
J_0&=&\sum_{m\geq1}\frac{-\Gamma(1/2)}{4m!\Gamma(m+1/2)}\big(\frac{z}{4}\big)^{m-1/2}\Gamma(m-1/2).\nonumber\\
&=&\sum_{m\geq1}\frac{-\sqrt{\pi}}{4m!(m-1/2)}\big(\frac{z}{4}\big)^{m-1/2}\nonumber\\
&=&-\sqrt{\pi}\frac{1}{16}\int _0^z\big(\frac{4}{\zeta}\big)^{3/2}\big(\exp(\zeta/4)-1\big)d\zeta\nonumber\\
&=&-\frac{\pi}{2}\Big\{\operatorname{erfi}\big(\frac{\sqrt{z}}{2}\big)+\frac{2}{\sqrt{z}}\big(1-\exp(z/4)\big)\Big\},\nonumber
\end{eqnarray}
where $\operatorname{erfi}()$ is the imaginary error function. Hence, we get
$$\kappa_t=\frac{\sqrt{\nu_t}}{T}\Big\{1-\sqrt{\pi}\Big[\frac{\sqrt{z}}{2}\operatorname{erfi}\big(\frac{\sqrt{z}}{2}\big)+1-\exp(z/4)\Big]+\sqrt{\frac{z}{\pi}}J_\infty\Big\}$$
We now procede to compute the other terms in $J_\infty$. There, we need to compute for $n\geq1$
\begin{eqnarray}
\int_0^\infty\frac{\exp(-\frac{y^2}{z})F_n(y)}{y^2}dy&=&\int_0^\infty\exp(-\frac{y^2}{z})\sum_{m\geq0}\frac{1}{m!\Gamma(m+2n+1/2)}\big(\frac{y}{2}\big)^{2m+2n}\frac{dy}{y^2}\nonumber\\
&=&\sum_{m\geq0}\frac{1}{4m!\Gamma(m+2n+1/2)}\int_0^\infty\exp(-\frac{y^2}{z})\big(\frac{y}{2}\big)^{2m+2n-2}dy\nonumber\\
&=&\sum_{m\geq0}\frac{1}{4m!\Gamma(m+2n+1/2)}\int_0^\infty\exp(-\frac{y^2}{z})\big(\frac{y^2}{4}\big)^{m+n-3/2}\frac{ydy}{2}\nonumber\\
&=&\sum_{m\geq0}\frac{1}{4m!\Gamma(m+2n+1/2)}\int_0^\infty\exp(-s)\big(\frac{sz}{4}\big)^{m+n-3/2}\big(\frac{z}{4}\big)ds\nonumber\\
&=&\sum_{m\geq0}\frac{1}{4m!\Gamma(m+2n+1/2)}\big(\frac{z}{4}\big)^{m+n-1/2}\int_0^\infty\exp(-s)s^{m+n-3/2}ds\nonumber\\
&=&\sum_{m\geq0}\frac{\Gamma(n+m-1/2)}{4m!\Gamma(m+2n+1/2)}\big(\frac{z}{4}\big)^{m+n-1/2}\nonumber\\
&=&\sum_{m\geq0}\frac{\Gamma(n-1/2)(n-1/2)^{(m)}}{4m!\Gamma(2n+1/2)(2n+1/2)^{(m)}}\big(\frac{z}{4}\big)^{m+n-1/2}\nonumber\\
&=&\frac{\Gamma(n-1/2)}{4\Gamma(2n+1/2)} \big(\frac{z}{4}\big)^{n-1/2}{}_1F_1(n-1/2;2n+1/2;z/4),\nonumber
\end{eqnarray}
where $_1F_1(a;b;z)$ is the confluent hypergeometric function. Inserting the exponential factors and the $a_n$'s, we arrive at
\begin{eqnarray}
J_\infty&=&-\sum_{n\geq1}a_n\exp\big(E_n(T+t_0-t)\big)\int_0^\infty\frac{\exp(-\frac{y^2}{z})F_n(y)}{y^2}dy\nonumber\\
&=&\sum_{n\geq1}\frac{(-1)^{n+1}\Gamma(n-1/2)^2}{4n!\Gamma(2n-1/2)}e^{E_n(T+t_0-t)}\big(\frac{z}{4}\big)^{n-1/2}{}_1F_1(n-1/2;2n+1/2;z/4)\nonumber
\end{eqnarray}
Note, that $J_0$ may also be expressed as a confluent hypergeometric function, as 
\begin{eqnarray}
J_0&=&\sum_{m\geq1}\frac{-\Gamma(1/2)\Gamma(m-1/2)}{4m!\Gamma(m+1/2)}\big(\frac{z}{4}\big)^{m-1/2}.\nonumber\\
&=&\frac{-\Gamma(1/2)}{4}\big(\frac{z}{4}\big)^{-1/2}\sum_{m\geq1}\frac{\Gamma(m-1/2)}{m!\Gamma(m+1/2)}\big(\frac{z}{4}\big)^{m}.\nonumber\\
&=&-\frac{\Gamma(-1/2)^2}{4\Gamma(-1/2)}\big(\frac{z}{4}\big)^{-1/2}\Big({}_1F_1(-1/2;+1/2;\frac{z}{4})-1\Big).\nonumber
\end{eqnarray}
We finally arrive at
\begin{gather}
\kappa_t=\frac{\sqrt{\nu_t}}{T}\sum_{n\geq0}b_ne^{E_n(T+t_0-t)}\zeta^{n}{}_1F_1(n-1/2;2n+1/2;\zeta)\label{sol1},
\end{gather}
where
\begin{eqnarray}
b_n&=&\frac{(-1)^{n+1}\Gamma(n-1/2)^2}{2\sqrt{\pi}n!\Gamma(2n-1/2)},\nonumber\\
E_n&=&\alpha^2n(2n-1),\nonumber\\
\nu_t&=&\int_{t_0}^{t_0+T}\sigma^2_tdt\text{   is the realized volatility up to $t$},\nonumber\\
\zeta&=&\frac{\sigma^2}{2\alpha^2\nu_t}=\frac{z}{4}\nonumber,\\
\sigma&=&\sigma_t \text{   is the volatility at time $t$ and ${}_1F_1(;;)$ are confluent hypergeometric functions.}\nonumber
\end{eqnarray}

The fair value of a volatility swap with $t_0\leq t\leq t_0+T$ is hence

\begin{gather}
p(t,T)\Big(\frac{\sqrt{\nu_t}}{T}\sum_{n\geq0}b_ne^{E_n(T+t_0-t)}\zeta^{n}{}_1F_1(n-1/2;2n+1/2;\zeta)-K\Big)\label{sol1b}.
\end{gather}

\subsection{Value at $t=t_0+T$.}

We replace $t$ with $T+t_0$. The formula \ref{sol1} simplifies to

\begin{gather}
\kappa_{t_0+T}=\frac{\sqrt{\nu_t}}{T}\Big\{\sum_{n\geq0}b_n\zeta^{n}{}_1F_1(n-1/2;2n+1/2;\zeta)\Big\}.
\end{gather}
Inserting the series expansion of the confluent hypergeometric function, we arrive at

\begin{eqnarray}
\kappa_{t_0+T}&=&\frac{\sqrt{\nu_t}}{T}\Big\{\sum_{n,m\geq0}\frac{(-1)^{n+1}\Gamma(n-1/2)^2}{2\sqrt{\pi}n!\Gamma(2n-1/2)}\frac{(n-1/2)^{(m)}}{(2n+1/2)^{(m)}m!}\zeta^{n+m}\Big\}\nonumber\\
&=&\frac{\sqrt{\nu_t}}{T}\Big\{\sum_{n,m\geq0}\frac{(-1)^{n+1}(2n-1/2)\Gamma(n+m-1/2)}{2\sqrt{\pi}n!\Gamma(2n+m+1/2)}\frac{\Gamma(n-1/2)}{m!}\zeta^{n+m}\Big\}\nonumber.
\end{eqnarray}
By regrouping the terms with same exponents, we get

\begin{eqnarray}
\kappa_{t_0+T}&=&\frac{\sqrt{\nu_t}}{T}\Big\{\sum_{s\geq0}\zeta^{s}\frac{\Gamma(s-1/2)}{2\sqrt{\pi}}\sum_{0\leq n\leq s}\frac{(-1)^{n+1}(2n-1/2)\Gamma(n-1/2)}{n!(s-n)!\Gamma(s+n+1/2)}\Big\}\nonumber.
\end{eqnarray}
The first term, i.e. the constant multiplying $\zeta^0$, is $1$. The second, with $s=1$ and $n=0,1$ is equal to $-(-1/2)\Gamma(-1/2)/\Gamma(3/2)+(3/2)\Gamma(1/2)/\Gamma(5/2)=-\Gamma(1/2)/\Gamma(3/2)+\Gamma(1/2)/\Gamma(3/2)=0$.\\
For $s=2$ and $n=0,1,2$, one obtains $-(-1/2)\Gamma(-1/2)/2\Gamma(5/2)+(3/2)\Gamma(1/2)/\Gamma(7/2)-(7/2)\Gamma(3/2)/2\Gamma(9/2)=-\Gamma(1/2)/2\Gamma(5/2)+3\Gamma(1/2)/5\Gamma(5/2)-\Gamma(3/2)/2\Gamma(7/2)=(-5\Gamma(1/2)+6\Gamma(1/2)-2\Gamma(3/2))/10\Gamma(5/2)=0$, and one may show by a recurrence relation, that all these sums equal zero. we thus arrive at
$$\kappa_{t_0+T}=\frac{\sqrt{\nu_t}}{T},$$
which is the expected result.\\
This is actually the boundary value for the functional calculus we are going to study in the next section

\section{Verification By Functional Calculus}

In the computation of the solution, we somewhat ignored the mathematical difficulties which could arise from the exchange of orders of integration, especially when taking expectations and integrating over a parameter was involved. In this paragraph, we investigate the solution (\ref{sol1}) in the light of Rama Cont's papers \cite{cont10} \& \cite{cont10b} on functional stochastic processes. As given by $(\ref{sol1})$, $\kappa_t$ is certainly well-defined. It can even be shown, that it is an analytical function. The question is, if it is truly part of the solution of the problem at hand. Observe, that
$$\kappa_t=\frac{1}{T}\mathbb{E}\big((\int_{t_0}^{t_0+T}\sigma^2_sds)^{1/2}\big\vert \sigma_t=\sigma\big)$$
is in fact an integral depending on the path chosen for $\sigma_t$.
As explained in Rama Cont's paper on functional calculus, we must have
\begin{equation}
D_t\kappa_t+\frac{1}{2}\alpha^2\sigma^2(\nabla_\sigma^v)^2\kappa_t=0,\label{cont1}
\end{equation}
where $D_t=\partial_t+\nabla_\sigma^h$ and $\nabla_\sigma^h$ is the horizontal derivative, while $\nabla_\sigma^v$ is the vertical derivative. In our case, $\nabla_\sigma^h\zeta=0$, $\nabla_\sigma^h\int_{t_0}^{t_0+T}\sigma_s^2ds=\sigma_t^2\theta(t-t_0)\theta(T+t_0-t)$, $\nabla_\sigma^v\int_{t_0}^{t_0+T}\sigma_s^2ds=0$, $\nabla_\sigma^v\zeta=2\zeta/\sigma_t$ and the usual chain rules apply. We also should have $\kappa_{t_0+T}=\sqrt{\nu_{t_0+T}}$, which was already verified.

Let us lighten up the notations and put $\tau:=T+t_0-t$ and $f_n(\zeta):={}_1F_1(n-1/2;2n+1/2;\zeta)$. $\kappa_t$ rewrites as:
$$\kappa_t=\frac{\sqrt{\nu_t}}{T}\sum_{n\geq0}b_ne^{E_n\tau}\zeta^nf_n(\zeta)$$

Computing derivatives, we get
\begin{eqnarray}
D_t\kappa_t&=&\partial_t\kappa_t+\nabla^h_\sigma\kappa_t=\sqrt{\nu_t}\Big\{-\sum_{n\geq0}b_nE_ne^{E_n\tau}\zeta^nf_n(\zeta)\Big\}\nonumber\\
&+&\sqrt{\nu_t}\alpha^2\zeta\Big\{\sum_{n\geq0}b_ne^{E_n\tau}\zeta^nf_n(\zeta)\Big\}\nonumber\\
&-&2\alpha^2\zeta^2\sqrt{\nu_t}\Big\{\sum_{n\geq0}b_ne^{E_n\tau}\zeta^{n-1}\Big[\zeta f'_n(\zeta)+nf_n(\zeta)\Big]\Big\}\nonumber\\
&=&2\alpha^2\sqrt{\nu_t}\Big\{
\sum_{n\geq0}\zeta^{n}f_n(\zeta)b_ne^{E_n\tau}\big(\frac{\zeta}{2}-\frac{E_n}{2\alpha^2}-\zeta n\big)\nonumber\\
&-&\sum_{n\geq0}\zeta^{n+2}f'_n(\zeta)b_ne^{E_n\tau}\Big\}\nonumber
\end{eqnarray}
on the other hand, 
\begin{eqnarray}
(\nabla_\sigma^v)^2\kappa_t&=&\sqrt{\nu_t}\nabla_\sigma^v\Big[\frac{\sigma}{\alpha^2\nu_t}\Big\{\sum_{n\geq0}b_ne^{E_n\tau}\zeta^{n-1}\Big[\zeta f'_n(\zeta)+nf_n(\zeta)\Big]\Big\}\Big]\nonumber\\
&=&\frac{1}{\alpha^2\sqrt{\nu_t}}\Big\{\sum_{n\geq0}b_ne^{E_n\tau}\zeta^{n-1}\Big[\zeta f'_n(\zeta)+nf_n(\zeta)\Big]\Big\}\nonumber\\
&+&\frac{\sigma^2}{\alpha^4(\nu_t)^{3/2}}\Big\{\sum_{n\geq0}b_ne^{E_n\tau}\zeta^{n-2}\Big[ 2n\zeta f'_n(\zeta)+n(n-1)f_n(\zeta)+\zeta^2f''_n(\zeta)\Big]\Big\}.\nonumber
\end{eqnarray}

Now, use the fact, that the hypergeometric functions are solution to the differential equation $$z{}_1F_1''(a;b;z)=(z-b){}_1F_1'(a;b;z)+a{}_1F_1(a;b;z).$$
In our case,
$$\zeta^2f''_n(\zeta)=\zeta(\zeta-2n-1/2)f'_n(\zeta)+\zeta(n-1/2)f_n(\zeta).$$
This yields
\begin{eqnarray}
\frac{\alpha^2\sigma^2}{2}(\nabla_\sigma^v)^2\kappa_t
&=&2\alpha^2\sqrt{\nu_t}\Big\{\sum_{n\geq0}b_ne^{E_n\tau}\zeta^{n}\Big[ \zeta^2 f'_n(\zeta)+(\zeta +n)(n-1/2)f_n(\zeta)\Big]\Big\}.\nonumber
\end{eqnarray}
By replacing $E_n$ with their values $\alpha^2n(2n-1)$, we see that the functional equation (\ref{cont1}) is satisfied, which justifies the equation (\ref{sol1b}) as the solution to the fair value for volatility swaps.

\section{Conclusion}    
      The fair value of a volatility swap under the SABR model and with $t\in[t_0,t_0+T]$ has the solution (\ref{sol1b}) . The relation     
$$\sum_{0\leq n\leq s}\frac{(-1)^{n+1}(2n-1/2)\Gamma(n-1/2)}{n!(s-n)!\Gamma(s+n+1/2)}=0$$was not yet proven but can be accepted by verification. As a next computation, one may look for the value of a volatility swap for $t<t_0$. The value for a volatility option is also a question that bears some interest and should be addressed in the future.
\index{conclusion}                 
\index{LaTeX}    

\end{document}